\documentstyle[sprocl,epsfig,wrapfig]{article}

\begin{document}

\pagestyle{empty}

\begin{flushleft}
\Large
{SAGA-HE-131-98
\hfill May 14, 1998}  \\
\end{flushleft}
 
\vspace{1.6cm}
 
\begin{center}
 
\LARGE{{\bf Spin asymmetries at RHIC}} \\
\vspace{0.2cm}

\LARGE{{\bf and polarized parton distributions}} \\

\vspace{1.1cm}
 
\LARGE
{S. Hino, M. Hirai, S. Kumano, and M. Miyama $^*$}         \\
 
\vspace{0.3cm}
  
\LARGE
{Department of Physics}         \\
 
\LARGE
{Saga University}      \\
 
\LARGE
{Saga 840-8502, Japan} \\

\vspace{1.0cm}
 
\LARGE
{Talk given at the 6th International Workshop on} \\

\vspace{0.1cm}

{Deep Inelastic Scattering and QCD} \\

\vspace{0.1cm}

{Brussels, Belgium, April 4 -- 8, 1998} \\

\vspace{0.05cm}

{(talk on April 7, 1998) }  \\
 
\end{center}
 
\vspace{0.7cm}

\vfill
 
\noindent
{\rule{6.0cm}{0.1mm}} \\
 
\vspace{-0.3cm}
\normalsize
\noindent
{* Email: 97sm16@edu.cc.saga-u.ac.jp, 
98td25@edu.cc.saga-u.ac.jp,} \\

\vspace{-0.5cm}
\noindent
{\ \ \ kumanos@cc.saga-u.ac.jp,
96td25@edu.cc.saga-u.ac.jp.} \\

\vspace{-0.5cm}
\noindent
{\ \ \ Information on their research is available 
 at http://www.cc.saga-u.ac.jp}  \\

\vspace{-0.5cm}
\noindent
{\ \ \ /saga-u/riko/physics/quantum1/structure.html.} \\

\vspace{+0.1cm}
\hfill
{\large to be published in proceedings by the World Scientific}

\vfill\eject
\setcounter{page}{1}
\pagestyle{plain}


\title{SPIN ASYMMETRIES AT RHIC \\ AND POLARIZED PARTON DISTRIBUTIONS}

\author{S. HINO, M. HIRAI, S. KUMANO, M. MIYAMA
\footnote{Information on their research is available at 
http://www.cc.saga-u.ac.jp/saga-u/riko \\/physics/quantum1/structure.html.}}

\address{Department of Physics, Saga University \\ 
         Saga 840-8502, Japan}

\maketitle\abstracts{
We discuss polarized parton distributions and their effects on spin
asymmetries at RHIC. In particular, transversity distributions and
transverse spin asymmetry are studied. 
First, we show the $Q^2$ evolution difference between a transversity
distribution and a corresponding longitudinally polarized distribution.
The difference could be an important test of perturbative QCD in
high-energy spin physics. Then, the transverse spin asymmetry $A_{TT}$ is
calculated with possible transversity distributions. Next, we study
antiquark flavor asymmetry $\Delta_{_T} \bar u/ \Delta_{_T} \bar d$
in the transversity distributions by using a simple model. Its effects
on the transverse spin asymmetry are also discussed. 
}

\section{Introduction}\label{sec:intro}

It is important to test the proton spin structure through transversely
polarized structure functions, particularly the leading-twist structure
function $h_1$. There are three major reasons for investigating the
transversity distribution $h_1$,  which is often denoted as $\Delta_{_T} q$
or $\delta q$. The first reason is to test our knowledge of high-energy spin
physics in another spin observable in addition to the longitudinally
polarized ones. The second is to study a relativistic aspect of nucleon
structure. Because nonrelativistic quark models predict the same transversity
distribution as the longitudinally polarized one, the difference could
reflect the relativistic aspect. The third could be more important. 
Because the transversity $Q^2$ evolution is very different from the
longitudinal one as shown in section 3, the difference is a good test of
perturbative QCD in spin physics.

The transversity distributions are expected to be measured in the
transversely polarized Drell-Yan process at RHIC. We should try to
understand the properties of $h_1$ before the experimental data are
taken. In this paper, we discuss the $Q^2$ evolution of the transversity 
distributions~\cite{am,h1nlo} and compare its results with those
of the longitudinally polarized ones.~\cite{hkm} Then, the transverse
spin asymmetry $A_{TT}$ is investigated in connection with the transversity
distributions.~\cite{hkm,hk} Next, a possible antiquark flavor asymmetry 
$\Delta_{_T} \bar u/ \Delta_{_T} \bar d$ is studied in a simple
quark model, and its effects on $A_{TT}$ are shown.

\section{$Q^2$ evolution equation for transversity distributions}
\label{sec:dglap}

The transversity distribution $\Delta_{_T} q$ can be expressed in the parton
model. It is given by the probability to find a quark with spin
polarized along the transverse spin of a polarized proton minus the
probability to find it polarized oppositely:
$\Delta_{_T} q = q_{\uparrow}-q_{\downarrow}$.
Its leading-order (LO) $Q^2$ evolution equation was derived in 1990,~\cite{am}
and the next-to-leading-order (NLO) form was completed in 1997.~\cite{h1nlo}

Because of the chiral-odd nature of the transversity distribution,
the gluon does not participate in the evolution equation. 
Therefore, the DGLAP evolution equation is very different from the ones
for the longitudinal evolution. It is simply given by a single
integrodifferential equation,
\begin{equation}
\frac{\partial}{\partial\ln Q^2} \, \Delta_{_T} q^\pm (x,Q^2) \, = 
\frac{\alpha_s (Q^2)}{2\pi} \,  
\int_x^1 \frac{dz}{z} \, \Delta_{_T} P_{q^\pm} (z) \, 
\Delta_{_T} q^\pm \left(\frac{x}{z}, Q^2 \right)
\ ,
\label{eqn:DGLAP1}
\end{equation}
where $\Delta_{_T} P_{q^\pm}$ is the splitting function for the
transversity distribution. 
The notation $q^\pm$ in the splitting function indicates
the $\Delta_{_T} q^+ = \Delta_{_T} q + \Delta_{_T} \bar q$ or 
    $\Delta_{_T} q^- = \Delta_{_T} q - \Delta_{_T} \bar q$
distribution type.
The $\alpha _s (Q^2)$ is the running coupling constant. 
The transversity NLO evolution is the same in the $MS$ and
$\overline{MS}$ schemes. Even though the distribution
may not be flavor nonsinglet, the evolution equation looks like
the ``usual" nonsinglet one without coupling to the gluon term.

Dividing the variables $x$ and $Q^2$ into small steps, we solve
the DGLAP integrodifferential equation by the Euler method in the
variable $Q^2$ and by the Simpson method in the variable $x$.~\cite{hkm}
Numerical results indicate that accuracy is better than
1\% in the region $10^{-5}<x<0.8$ if more than fifty $Q^2$ steps
and more than five hundred $x$ steps are taken. Our $Q^2$ evolution
program could be obtained upon email request.\footnote{
See http://www.cc.saga-u.ac.jp/saga-u/riko/physics/quantum1/program.html.}

\section{$Q^2$ evolution results}\label{sec:q2trans}

Because the transversity distributions themselves are not measured yet,
it takes time for finding their scaling violation. On the other hand,
the $Q^2$ dependence is important for predicting spin asymmetries.
In particular, the transversity evolution is very different from 
the longitudinal one as we show in this section. We discuss the evolution
results for the flavor singlet transversity distribution
$\Delta_{_T} q_s = \sum_i (\Delta_{_T} q_i + \Delta_{_T} \bar q_i)$.~\cite{hkm}
The evolution of the flavor asymmetric distribution
$\Delta_{_T} \bar u - \Delta_{_T} \bar d$
is discussed in section \ref{sec:asym}.
There is a problem in studying the transversity evolution in the sense
that the input distribution is not available at this stage. However,
it is known within quark models that the transversity distributions are
almost the same as the corresponding longitudinally polarized distributions.
Therefore, we may use a longitudinal distribution as an input transversity
distribution at small $Q^2$.

\begin{wrapfigure}{r}{0.46\textwidth}
   \begin{center}
   \epsfig{file=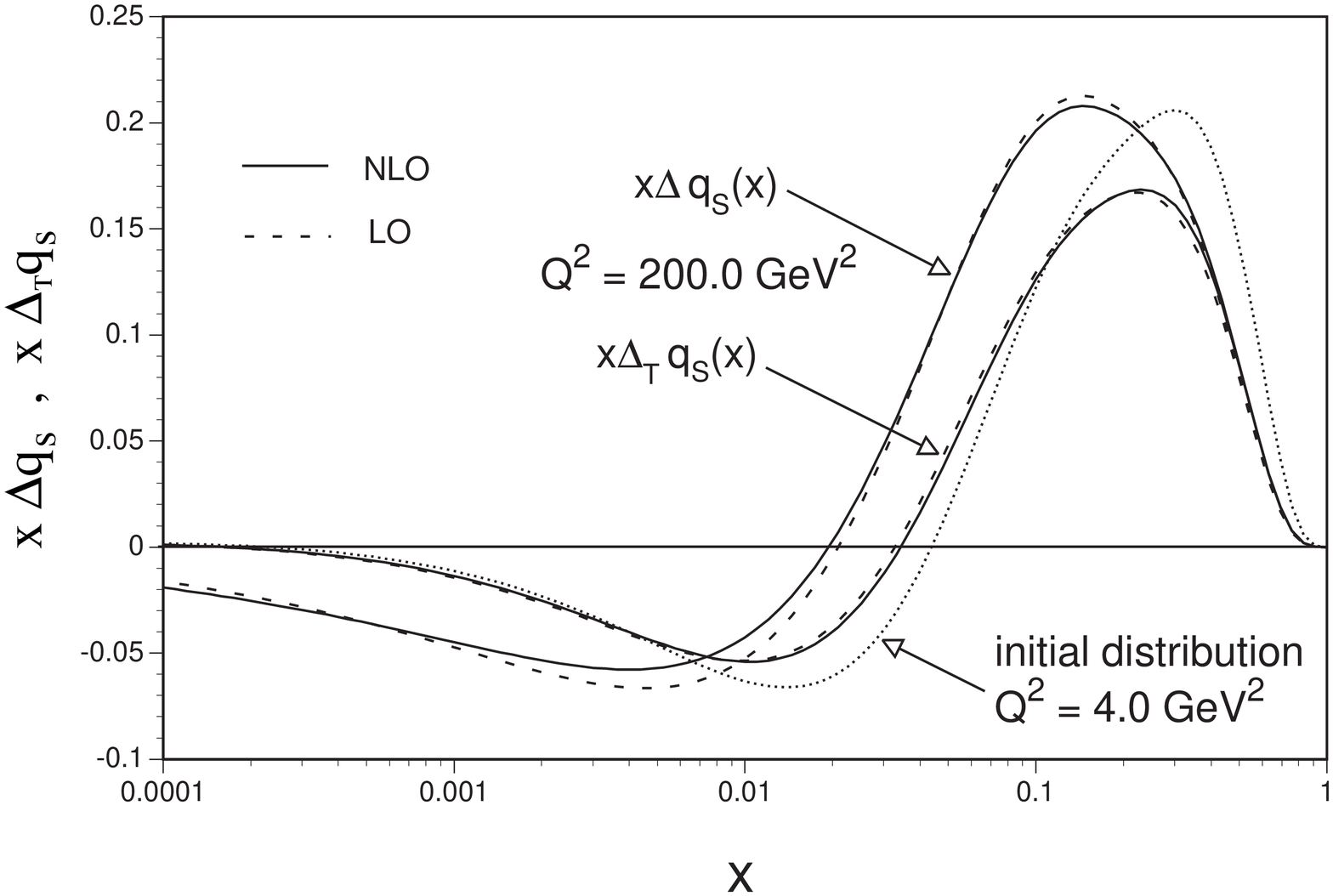,width=5.0cm}
   \end{center}
   \vspace{-0.4cm}
       \caption{\footnotesize 
          $Q^2$ evolution of the singlet transversity
          and longitudinally polarized distributions.}
       \label{fig:sing}
\end{wrapfigure}
The singlet evolution results are shown in Fig. \ref{fig:sing}.
The initial transversity and longitudinally polarized distributions are
assumed as the same GS-A distribution at $Q^2$=4 GeV$^2$, and they 
are shown by the dotted curve. It is evolved to the distributions
at $Q^2$=200 GeV$^2$ by the transverse or longitudinal evolution equation.
The LO and NLO evolution results are shown by the dashed and solid curves.
Because the evolution is from $Q^2$=4 GeV$^2$ to 200 GeV$^2$, the NLO
contributions are not so large. It is known that the NLO effects are
significant in the small $Q^2$ region, $Q^2<2$ GeV$^2$.
The transversity NLO effects increase the evolved distribution
at medium-large $x$ and also at small $x$ ($<$0.01), 
and they decrease the distribution in the intermediate
$x$ region ($0.01<x<0.1$). The transversity NLO effects are different
from the longitudinal NLO ones; however, it is more interesting to
find large differences between the evolved transversity and
longitudinally-polarized distributions.
For example, the evolved transversity distribution $\Delta_{_T} q_s$
is significantly smaller than the longitudinal one $\Delta q_s$
in the region $x\sim 0.1$. The magnitude of $\Delta_{_T} q_s$
itself is also smaller than that of $\Delta q_s$ at very small $x$
($<0.07$). Therefore, as we mentioned in the introduction, 
the study of the transversity distributions is important
for testing the perturbative aspect of QCD in spin physics.

\section{Antiquark flavor asymmetry and transverse spin asymmetry}
\label{sec:asym}

It is now well known that light antiquark distributions are not flavor
symmetric~\cite{sk} according to the NMC, NA51, and E866 experimental data.
In particular, the recent E866 Drell-Yan data revealed the $x$ dependence
of the $\bar u-\bar d$ distribution.~\cite{e866}
The mechanisms for producing the asymmetry are virtual meson clouds,
Pauli exclusion principle, and others. On the other hand, the antiquark
flavor asymmetry in the polarized distributions is not known at this stage
except for a few theoretical predictions. 
Because the polarized antiquark distributions are measured at RHIC,
it is important to investigate a possible asymmetric distribution.
In the following, we study the flavor asymmetry in the transversity
distributions.

\begin{wrapfigure}{r}{0.46\textwidth}
   \vspace{-0.0cm}
   \begin{center}
       \epsfig{file=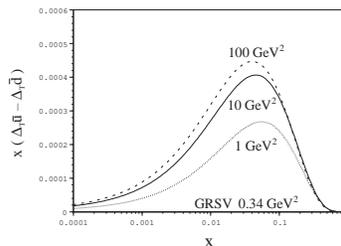,width=5.0cm}
   \end{center}
   \vspace{-0.4cm}
       \caption{\footnotesize
          $Q^2$ evolution of $x(\Delta_{_T}\bar u - \Delta_{_T}\bar d)$
          distribution. The initial distribution is 
          $\Delta_{_T}\bar u - \Delta_{_T}\bar d=0$ at $Q^2=0.34$ GeV$^2$.}
       \label{fig:grsv}
\end{wrapfigure}
First, we discuss the perturbative contributions.
They are expected to be small~\cite{sk} because there is no 
LO contribution. Due to the difference between the splitting functions
$\Delta_{_T} P_{q^\pm}$ in Eq. (\ref{eqn:DGLAP1}), there is a finite
perturbative  contribution to 
$\Delta_{_T} \bar u - \Delta_{_T} \bar d$.~\cite{mssv}
We choose the GRSV distributions at $Q^2$=0.34 GeV$^2$
as the initial ones although perturbative calculations may not be
valid in such a small $Q^2$ region. 
Despite the initial distributions are flavor symmetric, the NLO
evolution produces finite distributions in Fig. \ref{fig:grsv}. 
However, because the magnitude is rather small, 
the perturbative mechanism would not be
the major source for the flavor asymmetry. 

\begin{wrapfigure}{r}{0.46\textwidth}
   \vspace{-0.0cm}
   \begin{center}
       \epsfig{file=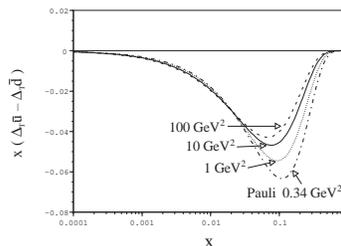,width=5.0cm}
   \end{center}
   \vspace{-0.4cm}
       \caption{\footnotesize 
          $Q^2$ evolution of $x(\Delta_{_T}\bar u - \Delta_{_T}\bar d)$
          distribution. The initial distribution is suggested by the Pauli
          exclusion model.}
       \label{fig:pauli}
\end{wrapfigure}
Many theoretical papers are written on the unpolarized flavor asymmetry
$\bar u-\bar d$. Although the meson-cloud mechanism is most successful
among the models, the polarized asymmetry is not well studied.
In order to estimate the order of magnitude of 
$\Delta_{_T} \bar u - \Delta_{_T} \bar d$ and its effects on
the transverse spin asymmetry, we use a simple picture
based on the Pauli exclusion principle.~\cite{excl}
Because the proton spin-up state is described in the SU(6) quark model as
$|p_+ \! \! > =$$ (1/\sqrt{6}) [ 2 | u_+ u_+ d_- \! \! >$$ 
                                 - | u_+ u_- d_+ \! \! >$$
                                 - | u_- u_+ d_+ \! \! > ]  ,$
we have each quark state probability as
$u_+ = 5/3$, $u_- = 1/3$, $d_+ = 1/3$, and $d_- = 2/3$.
These equations indicate that it is more difficult to create
the spin-up $u$ (spin-down $d$) quark than the spin-down $u$ (spin-up $d$)
according to the exclusion principle. Then, assuming that the exclusion
effect is the same as the unpolarized,
$(u_s^\downarrow - u_s^\uparrow)/(u_v^\uparrow - u_v^\downarrow) =
 (d_s - u_s)/(u_v - d_v)$ and a similar equation for 
$d_s^\uparrow - d_s^\downarrow$,
we have $\Delta_{(T)} \overline{u}= -0.13$ and 
$\Delta_{(T)} \overline{d}= +0.05$.~\cite{excl}
This exclusion model should be valid only at very small $Q^2$,
so that the GRSV parametrization is chosen in our analysis.
In order to estimate the distributions and the spin asymmetry $A_{TT}$,
the GRSV distributions are modified to have the first moments:
$\Delta_{T} \overline{u}= -0.13$ and
$\Delta_{T} \overline{d}= +0.05$.
The initial $\Delta_{_T} \bar u - \Delta_{_T} \bar d$ distribution
and its $Q^2$ evolution results are shown in Fig. \ref{fig:pauli}.
Because the polarization excess is larger in the $u$ quark, the exclusion
effect is dominated by the negative $\bar u$ quark polarization.

We have also calculated the transverse spin asymmetries at the RHIC
energy $\sqrt{s}=200$ GeV. In the flavor symmetric case, the Drell-Yan
spin asymmetry $A_{TT}$ is of the order of 0.5$\sim$1\% in the dimuon mass
region $100<M_{\mu\mu}^2<500$ GeV$^2$.
If the flavor asymmetry is taken into account, it increases to 1$\sim$2\%.
There is an indication of the $\Delta_{_T} \bar u  / \Delta_{_T} \bar d$
asymmetry in $A_{TT}$; however, the valence quark distributions have to
be fixed first in order to find $\Delta_{_T} \bar u$ and $\Delta_{_T} \bar d$.
Because $A_{TT}$ is rather small, we had better try other processes such
as $Z^0$ and jet production processes or semi-inclusive
muon scattering.~\cite{semi}
Because the transverse asymmetry $\Delta_{_T} \bar u  / \Delta_{_T} \bar d$
{\it cannot} be measured through the $W^\pm$ production processes,
we should think about possible measurements.~\cite{hkm}

\section{Summary}\label{sec:summary}
We have discussed the transversity distributions, in particular their
$Q^2$ evolution. Because the evolved transversity distribution is very
different from the longitudinally polarized one, the difference could be
an important test of perturbative QCD.
Next, we studied possible flavor asymmetric distributions
$\Delta_{_T} \bar u - \Delta_{_T} \bar d$ and their $Q^2$ evolution.
Because the perturbative QCD effects are rather small, we should
investigate nonperturbative mechanisms for creating the flavor
asymmetry. Calculated spin asymmetries $A_{TT}$ are rather small,
which suggests that we had better rely also on other measurements
for finding the accurate transversity distributions.


\end{document}